# Element-resolved ultrafast demagnetization rates in ferrimagnetic CoDy.

T. Ferte[1], N. Bergeard[1], L. Le Guyader[2*], M. Hehn[3], G. Malinowski[3], E. Terrier[1], E. Otero[4], K. Holldack[2], N. Pontius[2], and C. Boeglin[1]

[1]*Université de Strasbourg, CNRS, Institut de Physique et Chimie des Matériaux de Strasbourg, UMR 7504, F-67000 Strasbourg, France.*

[2]*Institut für Methoden und Instrumentierung der Forschung mit Synchrotronstrahlung Helmholtz-Zentrum Berlin für Materialien und Energie GmbH, Albert-Einstein-Str. 15, 12489 Berlin, Germany*

[3]*Institut Jean Lamour, Université Henri Poincaré, Nancy, France.*

[4]*Synchrotron SOLEIL, L'Orme des Merisiers, Saint-Aubin, 91192 Gif-sur-Yvette, France.*

[*]*Present Address: Spectroscopy & Coherent Scattering, European XFEL GmbH, Holzkoppel 4, 22869 Schenefeld, Germany*

**Abstract**:

Femtosecond laser induced ultrafast magnetization dynamics have been studied in multisublattice $Co_xDy_{1-x}$ alloys. By performing element and time-resolved X-ray spectroscopy, we distinguish the ultrafast quenching of Co3d and Dy4f magnetic order when the initial temperatures are below (T=150K) or above (T=270K) the temperature of magnetic compensation ($T_{comp}$). In accordance with former element-resolved investigations and theoretical calculations, we observe different dynamics for Co3d and Dy4f spins. In addition we observe that, for a given laser fluence, the demagnetization amplitudes and demagnetization times are not affected by the existence of a temperature of magnetic compensation. However, our experiment reveals a twofold increase of the ultrafast demagnetization rates for the Dy sublattice at low temperature. In parallel, we measure a constant demagnetization rate of the Co3d sublattice above and below $T_{comp}$. This intriguing difference between the Dy4f and Co3d sublattices calls for further theoretical and experimental investigations.

**I. INTRODUCTION**:

Ultrafast demagnetization induced by infrared femtosecond laser pulse in magnetic layers has been discovered 20 years ago [1]. Despite intensive theoretical [2, 3, 4, 5] and experimental [6, 7, 8] works, a comprehensive and unified model describing the ultrafast demagnetization and the transfer of the angular momentum at the sub-picosecond time-scale is still missing. A broad range of systems have drawn attention over those two decades, including metallic transition metals [1, 9, 10, 11], 4f lanthanides [12, 13, 14, 15, 16], semiconductors [17], half-metals [18] and oxides [19]. Since the discovery of all-optical switching (AOS) by femtosecond infrared (IR) laser pulses [20], laser induced magnetization dynamics in rare-earth - transition metal (RE-TM) alloys have received a legitimate interest [7, 14, 15, 21, 22, 23]. In these ferrimagnetic alloys, the TM 3d itinerant and the RE 4f localized magnetic moments are antiferromagnetically coupled through the RKKY indirect exchange mechanism [24, 25]. For some RE-TM concentrations, it is possible to define a temperature where the magnetization of the two sublattices compensate each other's, resulting in a zero net total magnetization. Such specific temperature is named the temperature of magnetic compensation $T_{comp}$ [26, 27, 28, 29, 30, 31, 32] and is a macroscopic property of the RE-TM system. If laser-induced spin dynamics would be driven by microscopic parameters only, then $T_{comp}$ would not be expected to affect the dynamics. This is however in contradiction with Medapalli *et al.* who studied laser-induced ultrafast demagnetization in RE-TM alloys as a function of temperature below and above $T_{comp}$ [33, 34]. They showed that the quenching of the 3d magnetic order is stronger when the initial temperature T of the alloy is below $T_{comp}$. However, they did not monitor the RE 4f spin dynamics [35]. During the same period, Lopez-Flores *et al.* reported different characteristic demagnetization times of the RE 4f sublattice for initial temperatures below and above $T_{comp}$ [14]. In this latter work, a faster demagnetization observed for $T < T_{comp}$ has been attributed to a critical slowing down of the excited 4f magnetic moments, near the Curie temperature ($T_C$), rather than an effect of $T_{comp}$ itself. Surprisingly, the authors did not observe such a critical slowing down for the 3d magnetic moments. Unfortunately, different RE-TM alloys with different magnetic properties (different RE elements, magnetic moments, anisotropies) were compared in this work. The interpretation given by Lopez-Flores *et al.* seems however in contradiction with recent theoretical results based on a Landau-Lifchitz-Bloch (LLB) model [36, 37, 38, 39, 40]. This model predicts that in RE-TM alloys, only the 3d spins undergo a critical slowing down in the vicinity of $T_C$. Therefore, a comprehensive element-resolved magnetization dynamics study across the compensation temperature is lacking.

In this publication, we address specifically the laser-induced ultrafast demagnetization dynamics in the vicinity of $T_{comp}$. We aim at extending the investigation of Medapalli *et al.* to the RE 4f demagnetization rates, in order to reveal ultrafast spin dynamics in the vicinity of $T_{comp}$ [33, 34, 41]. To do this we studied the ultrafast demagnetization rates of the Dy4f and Co3d magnetic sublattices in two CoDy alloys with identical static magnetic properties at the atomic scale (such as atomic magnetic moments and magnetic anisotropies) as verified in this work. Element-specific and time-resolved X-ray Magnetic Circular Dichroism (tr-XMCD) measurements were carried out at temperatures T below and above $T_{comp}$ for a given laser fluence allowing for an independent characterization of both sublattice demagnetization times and amplitudes [22, 14, 15, 42]. Our results reveal important differences with those from Medapalli *et al.* [33, 34]. We observed very close demagnetization amplitudes and characteristic demagnetization times for the Co sublattice for initial temperatures above and below $T_{comp}$ [33, 34]. Furthermore, we show that demagnetization amplitudes and characteristic demagnetization times are also similar in the case of the Dy sublattice. These results confirm that the temperature of magnetic compensation has no influence on these parameters. However, since the magnetization of the Dy sublattice is twice larger at $T<T_{comp}$ than at $T>T_{comp}$, our results show a twofold increase of the ultrafast demagnetization rate of the Dy 4f sublattice at low temperature. We show that in these CoDy alloy, none of the existing theoretical model is able to reproduce our experimental observations. Throughout this publication we aim at encouraging theoretical description of laser induced ultrafast dynamics taking into account the temperature of magnetic compensation.

**II. EXPERIMENT**

Time-resolved X-Ray Magnetic Circular Dichroism operated in the "*femtoslicing*" mode at BESSY II [34, 10, 35, 36] is the most suitable technique to extract the ultrafast magnetization dynamics in multisublattice alloys, combining the magnetic and elemental sensitivity [11, 12, 22, 14, 15, 42]. Femtosecond infrared laser pulses (with a wavelength of 800 nm, linearly polarized with a 60 fs pulse length) were used as the pump while the circularly polarized X-ray pulses with 100 fs duration were used as the probe [43]. The laser and X-ray spot size diameter at the sample location were approximately 500 μm and 150 μm respectively which ensure optimal spatial overlap. The laser fluence was set to a constant value of f = 7 mJ/cm² which ensured ~60% demagnetization at the Co $L_3$ edge [14, 15]. The repetition rate of the pump-probe experiment was 3 kHz, and the system relaxed toward its initial state between two subsequent laser excitations. The beamline energy was set to the Co $L_3$ or the Dy $M_5$ absorption edges. XMCD was monitored by recording the transmitted X-ray intensities for opposite magnetic fields (H= ± 0.55T) as a function of the delay between the pump and the probe. The magnetic field was applied in the direction defined by the X-ray beam.

The choice of the $Co_{1-x}Dy_x$ samples has been motivated by the limitations of the experimental set-up in terms of accessible temperatures and magnetic fields. Even by considering a very broad range of concentrations and RE elements, we were not able to determine a single TM-RE system with a moderate coercive field ($H_C$) below 0.55 T, below and above $T_{comp}$, in the limited temperature range of 80 K< $T_{comp}$ <320 K [27, 28]. Therefore, we measured two different alloys, $Co_{78}Dy_{22}$ and $Co_{80}Dy_{20}$ with $T_{comp}$ = 320K and 220K, respectively. It is worth noting that in the framework of the Mean Field Approximation (MFA), the microscopic magnetic properties such as the atomic magnetic moment ($\mu_i$) or the exchange coupling constant ($J_{ij}$) are considered independent of the composition and temperature [44]. Calculations based on MFA reproduced accurately the dependence of magnetic properties ($T_{comp}$, $T_{Curie}$, magnetization M and $H_c$) on composition and temperature [26, 27, 28, 45]. Therefore, although our samples display disparate macroscopic properties ($H_c$, M, $T_{comp}$), they can be considered similar at the microscopic scale. We confirmed this assumption by static X-ray transmission spectroscopy and XMCD measurements (see next section for details). The $Co_{78}Dy_{22}$ alloy displays $H_c$ < 0.55T at T < 150K while the $Co_{80}Dy_{20}$ alloy displays $H_c$ < 0.55T at T > 270K. The low magnetic coercive fields of the alloys ensure magnetic saturation under the applied +/-0.55 T magnetic field available on the FEMTOSPEX endstation. In addition, the experiment can be performed at temperatures below 300K as annealing effects have been observed for other sample composition at higher temperature. Following our static sample characterization we can safely compare the ultrafast dynamics measured for both samples. In the rest of the article, we define the following names S1 and S2 corresponding to respectively $Co_{80}Dy_{20}$ and $Co_{78}Dy_{22}$ for which the laser-induced dynamics was monitored at T> $T_{comp}$ and respectively T< $T_{comp}$ as illustrated in figure 1.

The CoDy alloys of 18 nm thickness were sputter-deposited on X-ray transparent 200 nm thick $Si_3N_4$ membranes, and capped with a 3 nm Ta layer to protect against oxidation. A heat sink and buffer layer composed of a 80 nm thick Ta/Cu multilayers was grown between the alloys and the membranes. This buffer layer limits the temperature increase due to laser DC-heating (estimated) to ~70K. The total X-ray transmission of the stacks in the vicinity of the Co and Dy absorption edges was about 50 %. The dependence of the alloys magnetization with temperature was measured by VSM-SQUID, to determine the temperature of magnetic compensation $T_{comp}$ of our CoDy alloys. This temperature is also an accurate indication of the effective composition since a tiny change in the concentration induces a large variation in $T_{comp}$ [27, 28].

The static XMCD spectroscopy was performed on the DEIMOS beamline at synchrotron SOLEIL in order to characterize the magnetic properties of the alloys. The measurements were performed in transmission and at normal incidence, i.e the same geometry as the one used in the pump-probe experiments at the "femtoslicing" beamline. The X-ray transmission spectra were recorded under a ± 2T magnetic field applied parallel to the X-ray beam for both left- and right-circular polarizations.

### III. EXPERIMENTAL RESULTS

**1. Magnetic properties**.

In figure 2a we show the X-ray transmission signals measured at the Dy $M_5$ edge at T=150 K and negative helicity for S1 and S2. At T=150K both alloys (S1 and S2) are characterized by the same multiplet structures and thus the same occupation of the Zeeman levels [46]. In addition, the static XMCD spectra recorded at the Dy $M_5$ edge are compared for both samples in figure 2b and show identical features at T=150K. The similarities of the multiplet structures and XMCD signals observed at T=150K ensures very close electronic and magnetic properties for S1 and S2. In figure 3a and 3b we show the transmission signal measured for positive and negative helicities, recorded at 150K at the Co $L_{2,3}$ and Dy $M_{4,5}$ edges for S1 and S2. The sum rules analysis, including the spin dipolar moments, was used in order to extract the magnetic moments per atoms $m_i$ of Co and Dy in both alloys (table 1) [47, 48, 49, 50]. Here, $m_i$ is defined as the sample averaged projection along the X-ray incidence, of the individual atomic magnetic moments $\mu_i$. These values, which are listed in table 1, show that the magnetic moments of Co and Dy are consistent with literature [46, 51, 52, 53] and they confirm that at T=150K, both alloys have the same magnetic moments within the error bars. We can thus safely compare the ultrafast magnetization dynamics below and above $T_{comp}$, comparing the data measured for S1 and S2.

The thermal dependences of the magnetic moments in Co and Dy and the coercive fields, have been analyzed in the temperature range used during the pump-probe experiments (150 –270 K). In Figure 4 we show the XMCD at Dy $M_5$ for both alloys at the temperatures of 150 K and 270 K. The inset shows the hysteresis curves extracted from the Dy $M_5$ XMCD data as a function of the applied field. The loops are square for both alloys S1 and S2 and confirm the limited coercive fields ($H_C < 0.55$ T). In Table 1 we can see that the Dy sublattice shows a lower magnetic moment at T=270K (m= -4.12 ± 0.21 $\mu_B$/at) than at T=150K (m=+6.64± 0.33 $\mu_B$/at) whereas the magnetic moment of Co does not change within the error bars. The quantitative magnetic moments measured for Co and Dy, as well as their thermal dependences, are consistent with experimental values obtained for similar alloys [46, 54]. At T=270K (compared to 150 K) the low Dy $M_5$ XMCD signal results from the large thermal fluctuations affecting the Zeemann levels. Therefore we can conclude that the temperature dependence of the Dy $M_5$ XMCD signal is not influenced by $T_{comp}$ [45]. Our quantitative element-resolved values of the magnetic moments will be used in the next section. They are important ingredients in order to derive the demagnetization rates (D) in each sublattice, representing the ability of the Co and Dy spins to transfer the angular momenta to other sub-systems as for instance the lattice.

**2. Ultrafast magnetization dynamics in CoDy alloys by tr-XMCD**:

The time-resolved XMCD signal recorded over short delay ranges (~ 4ps) at the Co $L_3$ and Dy $M_5$ edges are displayed in figure 5a and 5b, respectively. At negative delays t = $t^<$, the XMCD signals at the Dy $M_5$ edge are proportional to the static XMCD characterization performed at similar temperatures and given in table 1. Over the whole investigated delay range, the sign of the XMCD signal remains constant, excluding any laser induced magnetization reversal [22]. However, different demagnetization amplitudes as well as different dynamic of magnetization recovery are visible between Co and Dy at both temperatures. The transient XMCD signal at the Co $L_3$ edge shows relative demagnetization amplitude of ~60%, followed by the magnetization recovery. The transient XMCD signal at the Dy $M_5$ edge shows an almost completely quenched magnetization and no recovery. Furthermore, the Co magnetization reaches its minimum value (m(t*)) at a delay t*~0.5 ps for which Dy magnetization achieved only half of its total demagnetization. The Dy magnetization reaches its minimum value (m(t*)) at a delay t*~2 ps. Transient normalized XMCD at Co $L_3$ and Dy $M_5$ edges are shown in figure 6a and 6b in order to compare the element-resolved ultrafast demagnetization at temperatures T above and below $T_{comp}$ (T> $T_{comp}$ and T<$T_{comp}$) in each sub-system. The curves are fits to the data with a single exponential decay followed by an exponential recovery. Longer delay scans (not shown) have been performed in order to estimate the recovery times (table 2). The fit function is convoluted by a Gaussian function representing the time resolution of the experiments ~130 fs [10, 11]. The relative demagnetization amplitudes (A= $(m(t^<) - m(t^*)) / m(t^<)$) as well as the characteristic demagnetization times ($\tau$) and recovery times ($\tau_R$), extracted from the fit function, are reported in table 2. The ultrafast demagnetization rate (D) characterizes the ability of the spins of the Co and Dy sublattices to transfer the angular momentum and is calculated from our experimental data (table 2). We can define D as the ratio between the demagnetization amplitude (($m(t^<) - m(t^*)$)) and the characteristic demagnetization time $\tau$. It follows the equation [55-57]:

$D = (m(t^<) - m(t^*)) / \tau$ (in $\mu_B$/ps.at)

with m(t) the element-resolved magnetization (magnetic moment per atom given in $\mu_B$/at) at a delay t and $\tau$ the element-resolved characteristic demagnetization time.

For the Co sublattice, at T< $T_{comp}$ and T>$T_{comp}$, the laser induced ultrafast normalized demagnetization amplitudes A are 59±5 % and 65±4%, and the demagnetization times $\tau$ are 190±60 fs and 160±60 fs, respectively (Table2). Those values show that within the error bars, the extracted values D are almost similar at both temperatures (5.8±2.7 and 6.2±3.1 $\mu_B$/ps.at). For the Dy sublattice, at T< $T_{comp}$ and T>$T_{comp}$, the relative demagnetization amplitudes A are 80±9 % and 92±8 % and the demagnetization times $\tau$ are 610±70 fs and 630±60 fs, respectively. They also remain similar within our error bars. The extracted values D are however different at both temperatures (9.6±2.2 $\mu_B$/ps.at and 5.4±1.5 $\mu_B$/ps.at). This is a consequence of the difference between the Dy magnetization at T< $T_{comp}$ and T>$T_{comp}$ (Table 2). Our results show that at T< $T_{comp}$, D is nearly twice as large as at T>$T_{comp}$. Moreover, in spite of the strong 3d-4f indirect exchange coupling, the Co and Dy sublattices show different relative demagnetization amplitudes (average values ~60% and ~85% resp.) and different characteristic demagnetization times (average values ~170 fs and ~620 fs resp.) consistent with the various element- and time-resolved XMCD experiments reported in literature [14, 15, 22, 42, 58, 59].

**IV. DISCUSSION**

In a recent work, Medapalli *et al.* reported larger demagnetization amplitudes of the FeCo sublattice in FeCoGd alloys for initial temperatures below $T_{comp}$ [33, 34]. This is not observed for the Co sublattice in our CoDy alloys.

We show in this work that the conclusions extracted from FeCoGd alloys cannot be extended to our RE-TM alloys. It is worth noting that in FeCoGd alloys, experimental [7, 20-23] and theoretical [40, 41, 60- 63] results demonstrate that in addition to the laser induced demagnetization a particular Helicity-Independent All-Optical Switching (HI-AOS) of the spins is possible [64]. We note that the HI-AOS is observed only if the sample temperature is in the proximity of $T_{comp}$ [65]. Therefore, the enhanced demagnetization amplitudes in FeCoGd at $T< T_{comp}$ as reported by Medapalli *et al.* is probably a specificity of this alloy and a signature of HI-AOS [41, 64]. It has been shown that the demagnetization amplitudes upon laser excitation are fluence [66-68] and temperature dependent ($T-T_C$) [68]. Therefore, we can safely conclude that the similar demagnetization amplitudes which are observed in our study at $T<T_{comp}$ and $T>T_{comp}$ are a consequence of the close values of our experimental parameters, laser fluences and $|T-T_C| \sim 450\pm20$ K. Finally, our results further confirm that the macroscopic parameter $T_{comp}$ has no influence on the demagnetization amplitudes (($m(t^<) - m(t^*)$)).

Let us discuss the characteristic demagnetization times ($\tau$) extracted for Co and Dy (table 2). In a former publication, we have shown that larger values $\tau$ could be measured for Tb in $Co_{86}Tb_{14}$ ($\tau\sim500$ fs) compared to $Co_{74}Tb_{26}$ ($\tau\sim280$ fs) and we related them to the proximity of the temperature to $T_C$ ($T-T_C$) [14]. These results are also confirmed by Atxitia et al. [60] based on the LLB model, which shows that $\tau$ scales with $1/(T-T_C)$ when T is close to $T_C$.

In this work however, we use values of $|T-T_C|$ close to 450 K during the experiments at $T<T_{comp}$ and $T>T_{comp}$. Therefore, the close values of $\tau = 630 \pm 55$ fs and $\tau = 610 \pm 65$ fs measured for Dy are consistent with our previous conclusions where $\tau$ depends on $|T-T_C|$ [14]. In a recent work by Mentink *et al.* a phenomenological model was proposed in order to describe ultrafast spin dynamics in RE-TM alloys [70]. They show that in the temperature dominated regime, for which the electronic temperature ($T_e$) is far above $T_C$, $\tau$ is given by

$\tau = \mu / (2\alpha\gamma k_B T_e)$

with µ the atomic magnetic moment (independent from the thermal fluctuations), α the coupling constant with the heat bath, γ the gyromagnetic ratio and $k_B$ the Boltzmann's constant [36, 71]. Radu *et al.* applied this phenomenological model to establish a linear relation between τ and m with different RE-TM alloys [42]. In our work we can assume an equivalent elevation of the electronic temperature $T_e$ upon laser excitations for S1 and S2 which is guaranteed by the very same laser fluence used for both experiments. Assuming that the parameters $\mu_i$, $\alpha_i$ and $\gamma_i$ are similar for S1 and S2 we can stress that the phenomenological model from Mentink *et al.* supports our results and justifies the similarity of τ at $T<T_{comp}$ and $T>T_{comp}$.

Several other theoretical models such as the LLB model [39, 40, 60] and the microscopic 3 temperature model (m3TM) [3, 62] have related τ to microscopic magnetic properties of ferrimagnets. These models predict that the macroscopic parameter $T_{comp}$, has no influence on τ in line with our results. Our conclusions are further sustained by recent works from Rettig *et al.* considering the specific case of antiferromagnetic Ho layers [16]. They show that for Ho layers without finite magnetization, τ is similar to those reported for ferromagnetic Gd and Tb [12]. They evidenced that a zero net magnetization in antiferromagnetically coupled sublattices does not influence the demagnetization time.

In conclusion, we show that the $T_{comp}$ has no influence on the demagnetization amplitude and characteristic demagnetization times. This assumption is supported by phenomenological models assuming that these parameters are determined by the microscopic magnetic properties of our alloys while $T_{comp}$ is a macroscopic parameter.

So far, we have qualitatively discussed both features, defined by τ and by the amplitude of demagnetization for Co and Dy at temperatures T above and below $T_{comp}$. These quantities are used to define the demagnetization rates D characterizing the ability of the spin system to transfer the angular momenta to other sub-systems. The extracted demagnetization rates D for Co are 6.2±3.1 and 5.8±2.7 $\mu_B$/ps.at for T < $T_{comp}$ (sample S2) and T > $T_{comp}$ (sample S1) respectively (table 2). For the Dy sublattice, we observe a twofold increase of the demagnetization rate D at T < $T_{comp}$ (D= 9.6±2.2 $\mu_B$/ps.at) compared to T > $T_{comp}$ (D= 5.4±1.5 $\mu_B$/ps.at) (table 2). Comparing all those numbers, we observe very close values of D except for the Dy sublattice at T < $T_{comp}$.

In the framework of the LLB model, the demagnetization rates of both Fe and Gd sublattices in FeCoGd alloy scale with their respective magnetization (equation 10 in [60]). This model should in principal, reproduce our experimental observations. However, the LLB equation describes pure spin dynamics neglecting the contribution of the orbital moments to the magnetization. Therefore, the description of the laser induced demagnetization of the Dy sublattice in CoDy alloys may not be straight forwards by using the LLB model [73]. Recently, Donges et al. have derived the Landau-Lifshitz-Gilbert (LLG) equation to describe accurately the thermal magnetic properties of CoDy alloys [54]. Unfortunately, the authors have not extended their calculation to laser induced ultrafast dynamics yet. By comparing with literature, we can further support such temperature dependence of D for Dy and other the RE elements in RE-TM alloys. Radu et al. reported quantitative element-resolved investigation of laser induced ultrafast demagnetization in a $Co_{83}Dy_{17}$ alloy, measured at T > $T_{comp}$ [42]. We derived very close values of D for both Co and Dy sublattices: D=4.1±1.6 $\mu_B$/ps.at and D=4.2±1.3 $\mu_B$/ps.at respectively [42]. Furthermore, different values of D measured at different temperatures T > $T_{comp}$ and T < $T_{comp}$ can be extracted from the work of Lopez et al. in CoTb alloys [14]. The Tb sublattice shows following values: D= 6.12±3.2 $\mu_B$/ps.at for $Co_{86}Tb_{14}$ measured at T > $T_{comp}$ and D= 9.7±2.3 $\mu_B$/ps.at for $Co_{74}Tb_{26}$ measured at T < $T_{comp}$ [14]. The numbers and the increase of D at low temperature are coherent with the values extracted for the Dy sublattice in CoDy alloys suggesting that the effect is not a specificity of CoDy. All those values indicate a more efficient transfer of the angular momenta below $T_{comp}$ than above $T_{comp}$, for RE sublattices in RE-TM alloys.

Theoretical models have predicted that the proximity of $T_{comp}$ could have an influence on laser induced dynamics in RE-TM alloys. Gridnev et al. predicted recently that the exchange scattering should be more effective for T < $T_{comp}$ [41]. Barker et al. demonstrated that the magnon dispersions are dependent on the composition, and thus on the net magnetization, of the alloys [72]. Wiendholdt et al. predicted spin dynamics in the THz regime in the vicinity of $T_{comp}$ [32]. However, these models are focused on laser induced dynamics in FeCoGd alloys where the RE orbital magnetic moment is zero [33, 34]. Throughout this work, we aim at motivating additional experimental investigations and theoretical descriptions of laser induced dynamics in the vicinity of $T_{comp}$ in other RE-TM alloys than the FeCoGd alloys.

## V. CONCLUSIONS

We investigated the laser induced ultrafast magnetization dynamics with element selectivity in $Co_{78}Dy_{22}$ and $Co_{80}Dy_{20}$ alloys with initial temperatures below (T=150K) and above (T=270K) $T_{comp}$. We combined static and time-resolved XMCD spectroscopies in order to derive quantitative values for the demagnetization rates for each element and temperature. We demonstrated that the demagnetization amplitude and characteristic demagnetization times, determined by microscopic parameters, are not influenced by $T_{comp}$. The demagnetization rates are the same within the error bars for Co at both initial temperatures and for Dy at $T > T_{comp}$ while below $T_{comp}$ we observe a twofold increase in the demagnetization rate of Dy. It is not clear whether this enhanced demagnetization rate is induced by macroscopic properties of the alloys, such as $T_{comp}$. These measurements appeal for complementary experimental investigations at different temperatures, elements and laser fluences. We also hope these quantitative data will motivate further theoretical works in order to identify the role, or lack thereof, of the temperature of compensation on laser-induced ultrafast demagnetization in ferrimagnetic alloys.

**Acknowledgments:**

We are indebted to R. Mitzner and T. Kachel for help and support during the femtoslicing experiments. The authors are grateful for financial support received from the following agencies: The French "Agence National de la Recherche" via the projets ANR-11-LABX-0058_NIE, UMAMI ANR-15-CE24-0009 and the project EQUIPEX UNION: No. ANR-10-EQPX-52, the CNRS-PICS program, the European Union through FEMTOSPIN program and the EU Contract Integrated Infrastructure Initiative I3 in FP6 Project No. R II 3CT-2004-50600008. We acknowledge Synchrotron-SOLEIL for provision of synchrotron radiation facilities.

**Figures and Tables:**

| Sample | T (K) | Element | **m** spin (μ$_B$/at) | **m** orb (μ$_B$/at) | **m** (μ$_B$/at) |
|---|---|---|---|---|---|
| Co$_{80}$Dy$_{20}$ (S$_1$) | 150 | Co | -1.50 ± 0.08 | -0.19 ± 0.02 | -1.69 ± 0.10 |
| | | Dy | 2.98 ± 0.15 | 3.66 ± 0.18 | 6.64 ± 0.33 |
| | 270 | Co | 1.66 ± 0.08 | 0.22 ± 0.02 | 1.88 ± 0.10 |
| | | Dy | -1.89 ± 0.10 | -2.23 ± 0.11 | -4.12 ± 0.21 |
| Co$_{78}$Dy$_{22}$ (S$_2$) | 150 | Co | -1.35 ± 0.07 | -0.18 ± 0.02 | -1.53 ± 0.09 |
| | | Dy | 2.96 ± 0.15 | 3.59 ± 0.18 | 6.55 ± 0.33 |

**Table 1:** Static spin (**m** spin), orbital (**m** orb) and total (**m**) magnetic moments for Co and Dy obtained by applying the sum rules. The XMCD experiments were performed at the Co L$_{2,3}$ and Dy M$_{4,5}$ edges at T=150K for S$_1$ and S$_2$ and at T=270K for S$_1$. We show that both alloys have identical magnetic moments at T=150K within the error bars.

| Sample | T(K) | Element | τ (ps) | m(t$^<$) (μ$_B$/at) | A (%) | τ$_R$ (ps) | D = (m(t$^<$) − m(t$^*$)) / τ (μ$_B$/ ps.at) |
|---|---|---|---|---|---|---|---|
| Co$_{80}$Dy$_{20}$ (S$_1$) | 270 | Co L$_3$ | 0.19 ± 0.06 | 1.88 ± 0.10 | 59 ± 5 | 3 ± 1 | 5.8 ± 2.7 |
| | | Dy M$_5$ | 0.61 ± 0.07 | -4.12 ± 0.21 | 80 ± 9 | 21 ± 8 | -5.4 ± 1.5 |
| Co$_{78}$Dy$_{22}$ (S$_2$) | 150 | Co L$_3$ | 0.16 ± 0.06 | -1.53 ± 0.09 | 65 ± 4 | 3.5 ± 2 | -6.2 ± 3.1 |
| | | Dy M$_5$ | 0.63 ± 0.06 | 6.55 ± 0.33 | 92 ± 8 | > 100 | 9.6 ± 2.2 |

**Table 2:** Values of the relative demagnetization amplitude A =(m(t$^<$) − m(t$^*$))/ m(t$^<$), characteristic demagnetization time τ and recovery time τ$_R$, extracted from the fit function of the experimental demagnetization dynamics. D = =(m(t$^<$) − m(t$^*$))/ τ (in μ$_B$/ps.at) is calculated and represents the ultrafast demagnetization rates. The values m(t) are the element resolved magnetic moments, as extracted from the sum rules .(m(t$^<$) − m(t$^*$)) is the demagnetization amplitudes and τ the characteristic demagnetization times. The values are given for Co3d and Dy4f at T= 270K and 150K. The data sets are extracted from the fit parameters derived for the ultrafast dynamics recorded at each Co L$_3$ and Dy M$_5$ edges in S$_1$ at T=270K and S$_2$ at T=150K.

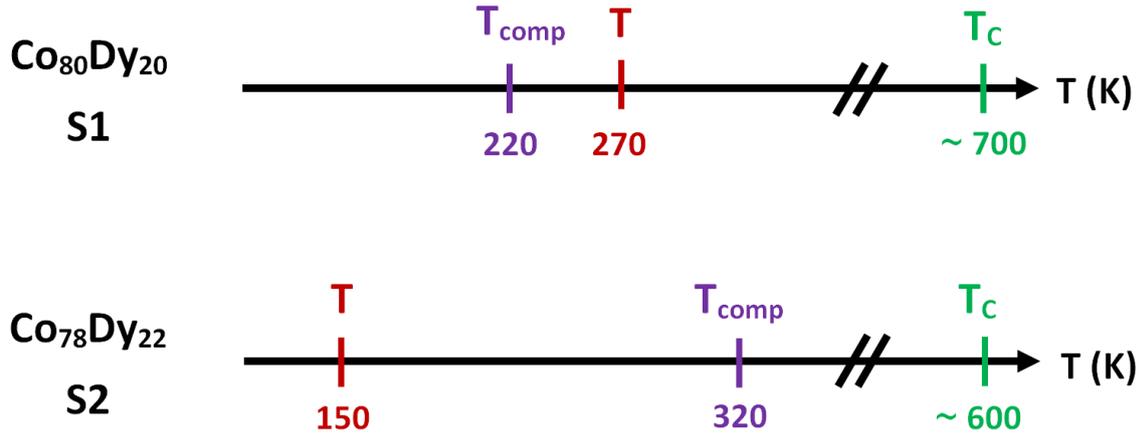

**Figure 1:** Sketch of the characteristic temperatures for $S_1$ and $S_2$. T is the temperature of the sample at negative delay (initial temperature) during the pump-probe experiment. $T_{comp}$ is the temperature of the magnetic compensation and $T_C$ the Curie temperature of the alloys. The values of $T_C$ are from Hansen *et al.* [27].

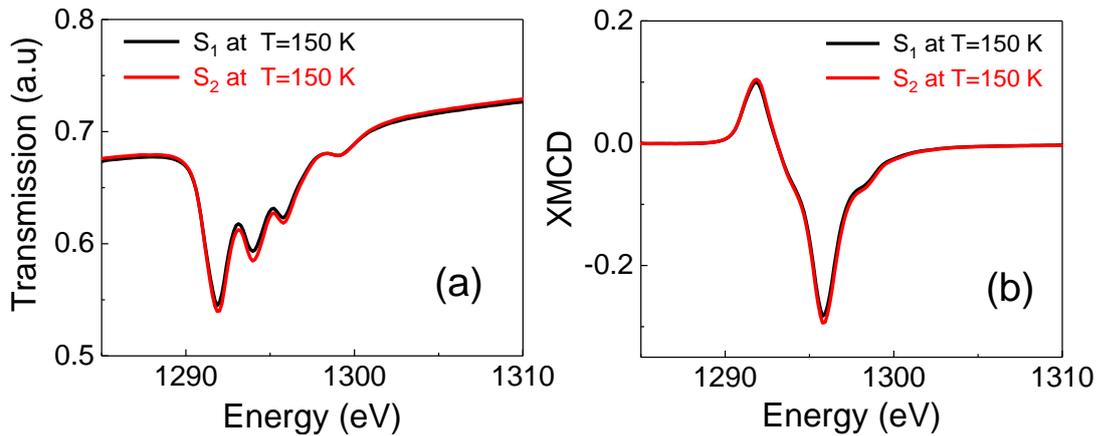

**Figure 2:** (a) X-ray transmission signals measured at negative helicity at the Dy $M_5$ edge for $Co_{80}Dy_{20}$ ($S_1$) (black line) and for $Co_{78}Dy_{22}$ ($S_2$) at T=150K (red line). At T= 150 K, the similarity of the multiplet structures show that both alloys can be assumed identical from the point of view of the electronic structure. (b) XMCD at the Dy $M_5$ edges for $Co_{80}Dy_{20}$ ($S_1$) (black line) and $Co_{78}Dy_{22}$ ($S_2$) at 150K (red line). The similarity of the XMCD amplitudes and shape at T=150K evidences that both alloys can be assumed identical from the point of view of the magnetic moments.

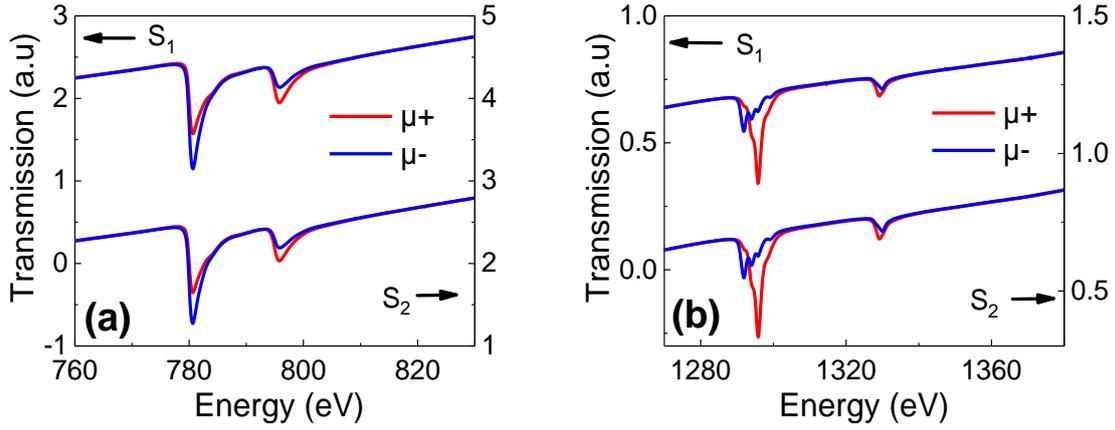

**Figure 3:** X-ray transmission signal measured at T=150K at positive and negative helicities at the (a) Co $L_{2,3}$ and (b) Dy $M_{4,5}$ edges for $Co_{80}Dy_{20}$ ($S_1$) and $Co_{78}Dy_{22}$ ($S_2$). The spectra are normalized such as $(\mu^+ + \mu^-)/2 = 1$ at the Co $L_3$ and Dy $M_5$ edges.

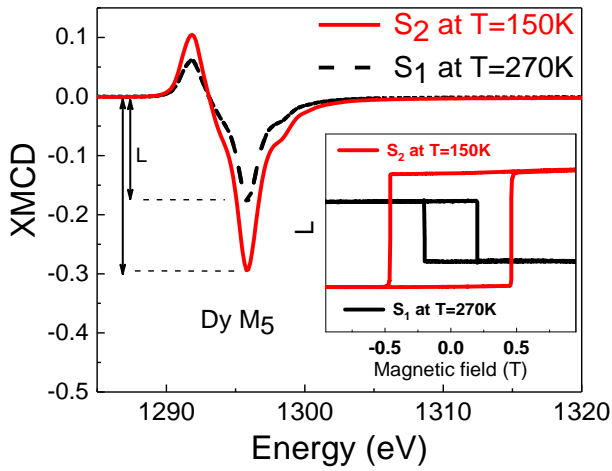

**Figure 4**: XMCD at the Dy $M_5$ edge for $S_1$ (black line) at T= 270K and $S_2$ at 150K (red line). The inset shows the magnetic hysteresis measured along the normal of the surface at the Dy $M_5$, defined by the XMCD amplitude L as illustrated. The hysteresis obtained for both alloys are superposed. They are measured at their temperatures used during the pump-probe experiments (270K and 150K). At these temperatures the alloys show identical anisotropies with low coercive fields ($H_C$) below 0.55T.

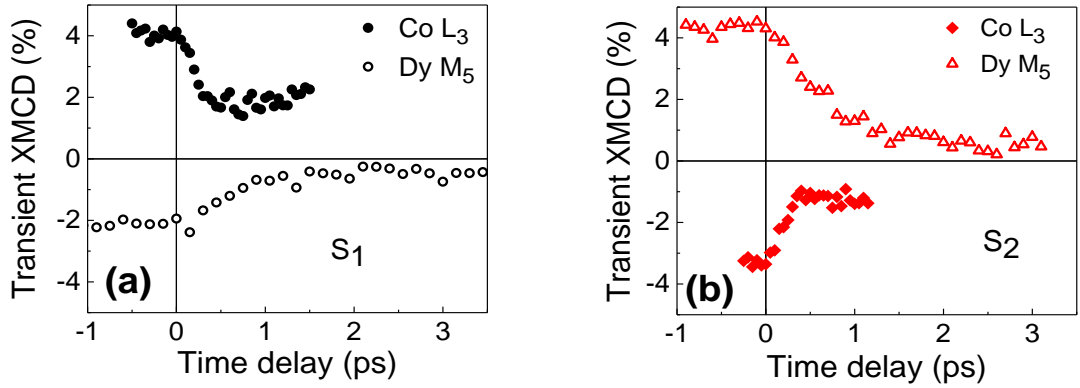

**Figure 5:** Transient XMCD at the Co $L_3$ (squares) and Dy $M_5$ (circles) edges for the $S_1$ (a) and the $S_2$ (b) alloys as a function of the delay. The laser fluence was 7 mJ/cm².

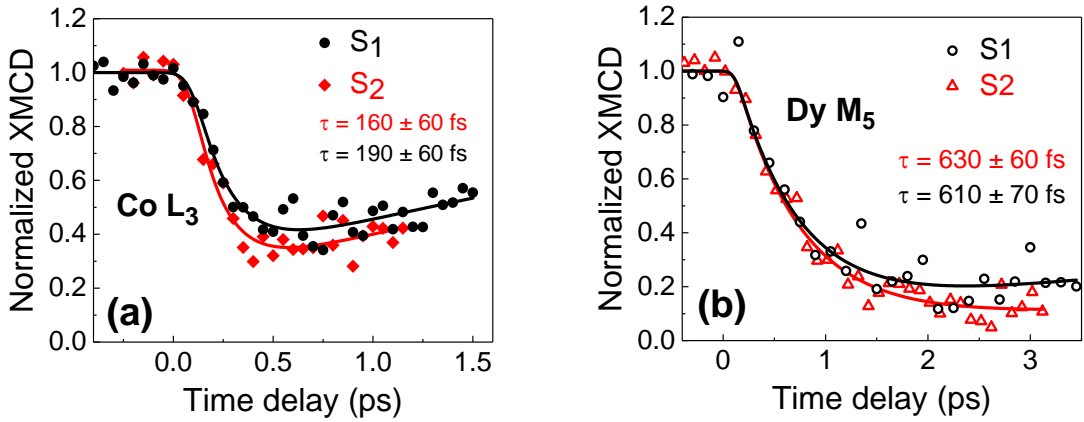

**Figure 6:** Transient normalized XMCD at the Co L3 (a) and Dy M5 (b) edges for the $S_1$ (black symbols) and the $S_2$ (red symbols) alloys as a function of the delay. The solid lines are fits to the data by a double exponential convoluted by a Gaussian function. The laser fluence was 7 mJ/cm².